\documentclass[12pt, a4paper]{article}
\usepackage{graphicx}
\usepackage{latexsym}
\usepackage{amsmath}
\usepackage{amssymb}
\usepackage{epsfig}

\begin{document}

\title{A Simple Proof of the Kochen-Specker Theorem on the Problem of Hidden Variables\footnote{Birthday contribution for Ruth Durrer, Geneva, January 25, 2008.}}

    \author{Norbert Straumann\\
        Institute for Theoretical Physics University of Zurich,\\
        CH--8057 Zurich, Switzerland}

\maketitle

\begin{abstract}
In this talk I present a simple derivation of an old result of Kochen and Specker, which is apparently unrelated to the famous work of Bell on hidden variables, but is presumably equally important. Kochen and Specker showed in 1967 that quantum mechanics cannot be embedded into a classical stochastic theory, provided the quantum theoretical probability distributions are reproduced and one additional highly desirable property is satisfied. This showed in a striking manner what were the difficulties in implementing the Einstein programme of a `complete' version of quantum mechanics.
\end{abstract}

\section{Introduction}

It is a great pleasure to be here at this happy occasion. Is it really true that Ruth is now fifty years old ? Congratulation and my best wishes.

As you heard from Martin, I was one of Ruth's teachers in Zurich. She was, as you can imagine, an extraordinary student: very interested in the subject and eager to really understand things. And she always solved the exercises. Obviously, a considerable fraction of her class mates usually copied Ruth's solutions. For that purpose they even expected from her, that she brought them sufficiently early.

I remember that during the breaks in one of the first courses she attended -- I believe it was electrodynamics -- Ruth regularly approached me and asked pertinent questions about a parallel course on quantum mechanics, given by one of my colleagues. She worried, for instance, how spin was introduced in that course, and found this for good reasons mysterious. We also discussed, of course, interpretational issues.- Well, I had only very few students like her.

With this background, I thought it would perhaps be fitting if I present something of pedagogical nature, that is at the same time of general interest, especially here in Geneva. I will talk about a very important old result of Kochen and Specker (KS), which is, unfortunately, not so widely known as the famous work of John Bell. Even people in the field were for a long time often not aware of the KS theorem. Loosely speaking, KS have shown that QM \emph{cannot} be embedded into a classical stochastic theory, provided that two very desirable conditions are assumed to be satisfied.

The original proof \cite{KS} of KS is very ingenious, but quite difficult. I show a crucial graph that plays an important role in the proof, and has become part of the cover sheet of an interesting book by M. Redhead \cite{Red}.

Ernst Specker was a young professor for mathematical logic when I was a student at ETH. I still see him from time to time. As a student I attended his course on foundations of analysis, and also some seminars on epistemology. I vividly remember Specker's talk on the problem of hidden variables in our joint theoretical seminar. Res Jost, who was a close friend of Specker, had invited him because he was well aware that the work with Kochen (who was then a postdoc of Specker) is interesting. At the time I did not grasp its full significance; that came much later.

Several years ago, when preparing the final lecture of my standard QM course, it suddenly occurred to me that a really simple proof of the KS-Theorem is possible. The evening before, S. Coleman had given a general physics colloquium in Zurich, in which he spent much time talking about the Greenberger-Horne-Zeilinger \cite{GHZ} version of a Bell type analysis. I then noticed -- and this is not a big achievement -- that the entangled states in their work can be used to prove the KS theorem at the blackboard in less than one hour. In my modest contribution I will mainly present this pedagogical exercise. I hope that some of you will enjoy this.

Before I start, let me remind you how the issue of hidden variables arose historically. Einstein, Schroedinger, and others were always hoping that physics would one day return to the reality concept of classical physics. At several occasions, Pauli emphasized that Einstein did not consider the concept of `determinism' to be as fundamental as it is frequently held to be. Einstein`s main concern was the radical revision of the concept of \emph{physical reality} by the Kopenhagen school. He maintained a realistic world view; he believed in a world of things existing as `real' entities. (Einstein was, of course, not a naive realist.)

A class of ``realistic theories'' is based on the idea of ``hidden variables'', by which one roughly means that QM is a kind of glorified statistical mechanics, which ignores some hidden microscopic degrees of freedom. This was at least the view of Einstein. I just quote one of his statements expressing this. In  the famous volume {\it Albert Einstein: Philosopher--Scientist} Einstein wrote \cite{Ein}:

\textit{``Assuming the success of efforts to accomplish a complete physical description, the statistical quantum theory would, within the framework of future physics, take an approximately analogous position to the statistical mechanics within classical mechanics. I am rather firmly convinced that the development of theoretical physics will be of this type; but the path will be lengthy and difficult.''}

With this background, let me now really begin.

\section{Formulation of the problem}

First I have to formulate the problem precisely, before stating the KS theorem. This requires some preparations and notation.

\subsection{Preliminaries}

Imagine some well-defined quantum mechanical system, for instance the spin-degrees of a spin-1 particle. Let $\mathcal{O}$ be a specified set of observables (e.g., a small finite number), represented by a set of self-adjoint operators of a Hilbert space $\mathcal{H}$. For $A\in\mathcal{O}$ we denote by $E^A(\cdot)$ the (projection-valued) spectral measure of $A$. We have the spectral decomposition
\begin{equation}
A=\int_{\sigma(A)} \lambda\;dE^A(\lambda),~~ \sigma(A): \textrm{spectrum of}~ A.
\end{equation}
More generally, the operator $u(A)$ for a Borel function $u:\mathbb{R}\rightarrow\mathbb{R}$ has the representation
\begin{equation}
u(A)=\int_{\sigma(A)} u(\lambda)\;dE^A(\lambda);
\end{equation}
in particular, $1_\Delta(A)=E^A(\Delta)$. We have the rule (as part of the `symbolic calculus')
\begin{equation}
 u_1(A)u_2(A)=(u_1\cdot u_2)(A).
\end{equation}
Furthermore, $E^{u(A)}$ and $E^A$ are naturally related by the push-forward:
\begin{equation}
E^{u(A)}(\Delta)=E^A(u^{-1}(\Delta)):~~E^{u(A)}=u_\ast E^A.
\end{equation}

Beside the observables $\mathcal{O}$ we also have to specify a set $\mathcal{S}$ of pure states (unit rays). (We could, more generally, choose mixed states, but this is not essential.) To each pair $[\psi]\in\mathcal{S},~A\in\mathcal{O}$ belongs the probability measure
\begin{equation}
w^A_{[\psi]}(\Delta)=(\psi,E^A(\Delta)\psi),
\end{equation}
and according to Born this is the distribution of $A$ in the state $[\psi]$. Note that (4) implies
\begin{equation}
w^{u(A)}_{[\psi]}=u_\ast w^A_{[\psi]}.
\end{equation}
(Remember that the induced transformation of a measure $\mu$ is given by $(u_\ast\mu)(\Delta)=\mu(u^{-1}(\Delta))$.)

\subsection{Hidden variables}

What do we mean when we ask whether this quantum description can be embedded into a classical theory, or be replaced by a theory with hidden variables? A minimal requirement, most people would agree on, is the following: There should exist a measurable space $(\Omega,\mathcal{F})~ (\mathcal{F}:\sigma$-algebra of measurable subsets) and two maps
\begin{equation}
\mathcal{O}\ni A \mapsto f_A: \Omega \rightarrow \mathbb{R}~~(\textrm{measurable} ),
\end{equation}
\begin{equation}
\mathcal{S}\ni[\psi]\mapsto \rho_{[\psi]}: \textrm{prob. measure on}~ (\Omega,\mathcal{F}),
\end{equation}
such that the probability distributions are reproduced:
\begin{equation}
 w^A_{[\psi]}(\Delta)=\rho_{[\psi]}(f^{-1}_A(\Delta)):~~~~\textrm{(\textbf{KS1})}.
\end{equation}
The first map assigns `values' to observables. The right hand side of (KS1) is the classical distribution of $f_A$ in the state $\rho_{[\psi]}$ (as in statistical mechanics).
In particular, the expectation values have to agree:
\begin{equation}
(\psi,A\psi)=\int_{\sigma(A)}\lambda\; dw^A_{[\psi]}(\lambda)=\int_\Omega f_A(\omega)\;d\rho_{[\psi]}(\omega).
\end{equation}
(In the last equality sign we have used a standard transformation formula for integrals.)

As long as not more is required, hidden variables in this sense can always be introduced. Following KS, this can be shown with the following abstract construction (which is quite natural from the point of view of probability theory): We choose
\begin{equation}
\Omega=\mathbb{R}^{\mathcal{O}}=\{\omega|\omega:\mathcal{O}\rightarrow\mathbb{R}\},~~\mathcal{F}=\mathcal{B}^{\mathcal{O}}
\end{equation}
($\mathcal{B}:\sigma-$algebra of $\mathbb{R}$), and the two maps $f_A,~\rho_{[\psi]}$ as
\begin{equation}
f_A(\omega)=\omega(A)~~(\textrm{canonical proj.}),~~ \rho_{[\psi]}=\bigotimes_{A\in\mathcal{O}}w^A_{[\psi]}~~
(\textrm{product measure}).
\end{equation}
(KS1) is indeed satisfied:
\[\rho_\psi(f_A^{-1}(\Delta))=\rho_\psi(\{\omega|\underbrace{f_A(\omega)\in\Delta}_{\omega(A)\in\Delta}\})=w^A_\psi(\Delta).\]

Note that in this construction the $f_A$ are measurable functions, which can be interpreted as random variables, but they are obviously \emph{independent}. However, observables in a physically interesting theory are usually not independent, and therefore we have to add some additional condition, in order to exclude this somewhat trivial construction.

In order to motivate the additional condition imposed by KS, we note that in any physical theory a function of an observable has to be defined such that the probability distributions for any state are related by push-forward, because this just reflects how functions of an observable are measured. This is indicated by the vertical arrows of the following diagram:
\vspace{0.5cm}

$\hspace{2cm}\underline{qm}~~~~~~~~~~~~~~~~~~~~~~~~~~~~~~~~\underline{cl}$

$\hspace{2cm}w^A_{[\psi]}~~~~~~~~~~~~~~~~~~~~~~~~~~~~~~w^a_\mu ~~~[w^a_\mu(\Delta)=\mu(a^{-1}(\Delta))]$

$\hspace{2cm}u_\ast$
$\hspace{-0.3cm}\Bigg\downarrow~~~~~~~~~~~~~~~~~~~~~~~~~~~~~~~\Bigg\downarrow\hspace{0cm}u_\ast$

$\hspace{2cm}w^{u(A)}_{[\psi]}~~~~~~~~~~~~~~~~~~~~~~~~~~~~w^{u(a)}_\mu$
\vspace{0.5cm}

(KS1) postulates that the upper distributions are equal if $a=f_A,~\mu=\rho_{[\psi]}$. The images under $u$ are then also equal, and thus (KS1) for the lower line then implies that $f_{u(A)}$ and $u(f_A)$ have the same distributions. It is very natural to require that the two functions are actually equal:
\begin{equation}
f_{u(A)}=u(f_A):~~~~(\textrm{\textbf{KS2}}).
\end{equation}
This condition has, as we shall see, far reaching consequences. It should be regarded as an alternative to Bell's locality assumption.

The following remark will be important. If $A_1,A_2\in\mathcal{O}$ are \emph{commuting}, then
\begin{equation}
f_{A_1A_2}=f_{A_1}\cdot f_{A_2},~~f_{A_1+A_2}=f_{A_1}+f_{A_2}.
\end{equation}
\textit{Proof}: The commuting selfadjoint operators $A_1,A_2$ can, by a theorem of von Neumann, be represented as real functions of a single selfadjoint operator $A$: $A_1=u_1(A),A_2=u_2(A).$ Then, by Eq. (3)
\begin{eqnarray*}
A_1A_2 &=& u_1(A)u_2(A)=(u_1\cdot u_2)(A);\\
f_{A_1A_2} &=& f_{(u_1\cdot u_2)(A)}\stackrel{\mathrm{KS2}}{=}(u_1 \cdot u_2)\circ f_A \\ &=&(u_1\circ f_A)\cdot(u_2\circ f_A)
\stackrel{\mathrm{KS2}}{=}f_{u_1(A)}\cdot f_{u_2(A)}=f_{A_1}\cdot f_{A_2}.
\end{eqnarray*}
The additivity follows similarly.

In Bell's work, the product rule in Eq. (14) follows for separated situations from his (Einstein's) locality assumption, however, not only for compatible observables.

\section{The KS theorem and its proof}

Now, we can formulate the main result of Kochen and Specker:
\vspace{0.3cm}

\textbf{THEOREM}. \textit{If dim$\mathcal{H}>2$, an embedding, satisfying (KS1) and (KS2), is ``in general'' -- for a large class of relevant examples -- not possible.}
\vspace{0.3cm}

\textit{Remarks}. 1) For the spin-degrees of freedom for a spin-1/2 particle such an embedding is possible, as was shown by J. Bell, as well as by KS (see Appendix).

2) We did not say precisely what we mean by `in general', because for that we would have to exclude cases for which the sets $\mathcal{O},\mathcal{S}$ would be too simple. What really matters, however, is that the statement is true for instance for most of atomic physics systems.

3) In the nineties several simplified proofs have been found (see, e.g., \cite{Per}), but the one I will now present is by far (?) the simplest.

\textit{Proof of the KS theorem}. Consider the quantum mechanical system describing the spin degrees of freedom of three spin-1/2 particles. The Hilbert space is $\mathcal{H}=\mathbb{C}^2\otimes \mathbb{C}^2 \otimes \mathbb{C}^2$, and important observables will be ($\sigma_x,\sigma_y,\sigma_z$ are the Pauli matices):
\begin{eqnarray}
 A_1 &=& \sigma_x\otimes 1\otimes 1,~~ A_2=1\otimes\sigma_x \otimes 1,~~ A_3=1\otimes 1\otimes \sigma_x,\\
 B_1 &=& \sigma_y\otimes 1\otimes 1,~~ B_2=1\otimes\sigma_y \otimes 1,~~ B_3=1\otimes 1\otimes \sigma_y.
\end{eqnarray}
Of special interest are the products
\begin{equation}
Q_1=A_1B_2B_3~ (=\sigma_x\otimes\sigma_y\otimes\sigma_y),~ Q_2=B_1A_2B_3,~ Q_3=B_1B_2A_3.
\end{equation}
One readily sees that all the $Q$'s commute and that $Q_j^2=1~(j=1,2,3)$.

Let $\chi_\uparrow,\chi_\downarrow$ be the canonical basis of $\mathbb{C}^2$. Consider the symmetric state
\begin{equation}
\Psi=\frac{1}{\sqrt{2}}[\chi_{\uparrow}\otimes \chi_{\uparrow}\otimes\chi_{\uparrow}-\chi_{\downarrow}\otimes \chi_{\downarrow}\otimes\chi_{\downarrow}].
\end{equation}
One easily sees that
\begin{equation}
Q_j\Psi=\Psi ~(j=1,2,3).
\end{equation}
Thus
\begin{equation}
\langle A_1B_2B_3\rangle_{\Psi}=1,~\textrm{etc. (strict correlations)}.
\end{equation}

Assume now the existence of an embedding, satisfying (KS1) and (KS2). Below we show that this implies a mathematical contradiction. Let
\[Q_j\mapsto f_{Q_j}\equiv q_j,~~ A_j \mapsto a_j,~~ B_j\mapsto b_j,~~ \Psi \mapsto \rho_{[\Psi]}.\]
Since the classical and quantum mechanical expectation values have to agree, we have
\[\langle Q_j\rangle_{\Psi}=1 \stackrel{\mathrm{KS1}}{=}\int_\Omega q_j(\omega)\;d\rho_\Psi(\omega)~~ (j=1,2,3).\]
Now (KS2) implies $q_j^2=1$ (the unit operator goes into 1, as a result of the product rule in (14)), hence $q_j(\omega)=\pm 1$. Together, we conclude that
\begin{equation}
 q_j(\omega)=1,~a.e.~~\Rightarrow q_1q_2q_3=1,~~ a.e.\;.
\end{equation}
Since the three factors in each $Q_j$ commute, we have by (14)
\[q_1=a_1b_2b_3,~q_2=b_1a_2b_3,~q_3=b_1b_2a_3.\]
Furthermore, $A_j^2=1, ~B_j^2=1$, thus $a_j(\omega)=\pm 1,~~b_j(\omega)=\pm 1$ and therefore we obtain from (21)
\begin{equation}
a_1a_2a_3=1,~~a.e.\;.
\end{equation}
(The $b$-factors appear quadratically.)

Now, the contradiction arises as follows. First one notes the operator identity
\begin{equation}
Q_1Q_2Q_3=-A_1A_2A_3 ,
\end{equation}
whence
\begin{equation}
(\Psi,A_1A_2A_3\Psi)=-1.
\end{equation}
Since the three factors in (24) commute, the image of $A_1A_2A_3$ is $a_1a_2a_3$, and (KS1) implies
\[\int_\Omega a_1(\omega)a_2(\omega)a_3(\omega)\;d\rho_\Psi(\omega)=-1,\]
hence
\begin{equation}
a_1a_2a_3=-1,~~a.e.\;.
\end{equation}
This is in ``maximal'' contradiction to (22).

\section{Concluding remarks}

I find the result of KS entirely satisfactory. It demonstrates clearly that there is no way back to classical reality. Notice that we considered only a small finite set of observables ($A_j,B_j,Q_j$), and we used only one state. It is physically important that we do not have to make assumptions on an infinite number of quantum mechanical propositions. Note also: If our system is a subsystem of a larger one (e.g., a system of $\geq$ 3 electrons in atomic physics), the larger one can also not be embedded in the sense of KS.

Of course, as with any theorem, there is the possibility to weaken the assumptions and hope for ways to avoid the far reaching conclusions. One suggestion (van Fraassen) goes as follows:

Maintain the idea of hidden variables (in the sense I used it). Keep, however, the map $A\rightarrow f_A$ only for \emph{maximal} operators $A$ (no degeneracies), but allow the possibility that for nonmaximal operators there may be several observables:

Suppose that for a nonmaximal $A$ we have the representations $A=u(B),~A=v(C),~B,C$ maximal, $[B,C]\neq 0$. If $B\rightarrow f_B,~C\rightarrow f_C$, we may associate to $A$ the functions on $\Omega$:
\[a_B=u(f_B),~~~a_C=v(f_C).\] There is no reason that the two agree. To make a choice requires a ``context'', and this led to the concept of \emph{``ontological contextuality''}. This goes in the direction of Bohr's mutual exclusiveness of experimental arrangements. The interesting question then arises, whether there is a connection of this with \emph{nonlocality.}

For more on these subtle issues I refer once more to Redhead's book \cite{Red}.

With good reasons one may consider the problem of hidden variables as irrelevant and mainly psychologically interesting. Thanks to KS (and, of course, J. Bell) we can, however, tell our students something definite.

\section{Appendix: Classical model for spin-1/2}

For the spin-1/2 degrees of freedom there is a classical model in the sense of KS.

I repeat part of the construction. The phase space $\Omega$ is the 2-sphere $S^2$ (with the Borel $\sigma$-algebra). The map $f_A$ is defined as follows:

Let $\lambda_1,\lambda_2$ be the eigenvalues of a hermitian matrix of $A,~A\psi_i=\lambda_i\psi_i$. If $\lambda_1\neq \lambda_2$ we associate to $A$ the traceless hermitian matrix (spin matrix)
\begin{equation}
\sigma(A)=\frac{2}{\lambda_1-\lambda_2}A-\frac{\lambda_1+\lambda_2}{\lambda_1-\lambda_2}1_2,
\end{equation}
for which $\sigma(A)\psi_{1,2}=\pm\psi_{1,2}$. Hence, $\sigma(A)$ can be written as $\sigma(A)=\vec{\sigma}\cdot\vec{x},~\vec{x}\in S^2.$ We call $S_A^+$ the hemisphere of $S^2$ with North Pole at $\vec{x}$. Then we set
\begin{equation}
f_A(p):=\left\{ \begin{array}{r@{\quad:\quad}l} \lambda_1 & p\in S_A^+,\\ \lambda_2 & \textrm{otherwise.}\end{array}\right.
\end{equation}
If the eigenvalues of $A$ are equal, so that $A=\lambda 1_2$, then we set $f_A(p)\equiv\lambda$. It is easy to show that this construction satisfies (KS2).

The probability measure on $S^2$ is also quite simple. We choose $\rho_\psi$ absolutely continuous with respect to the natural (rotational invariant) measure on $S^2$ (=\;$\sin\vartheta\; d\vartheta\; d\varphi$). The corresponding probability density $m_\psi(p)$ is constructed as follows. If $\psi=\chi_\uparrow$, we take
\[m_\psi=\left\{ \begin{array}{r@{\quad:\quad}l} \frac{1}{\pi}\cos\vartheta & 0\leq\vartheta\leq\pi \\ 0 & \textrm{otherwise,}\end{array}\right.\]
where $\vartheta$ is the polar angle of $p$. The generalization to arbitrary states is dictated by the requirement of rotational invariance. It then takes some calculations to verify that (KS1) is fulfilled. Actually, one can derive the formula for $m_\psi$ by imposing (KS1) (for details, see \cite{KS}).


\begin{thebibliography}{10}

\bibitem{KS}
S. Kochen and E. Specker, \textit{The Problem of Hidden Variables in Quantum Mechanics}, Journal of Mathematics and Mechanics \textbf{18}, 59-87 (1967).
\bibitem{Red}
M. Redhead, Incompleteness, Nonlocality, and Realism, A Prolegomenon to the Philosophy of Quantum Mechanics, Clarendon Press, Oxford (1987).
\bibitem{GHZ}
D. Greenberger, M. Horn and A. Zeilinger, in \textit{Bell's Theorem, Quantum Theory, and Conceptions of the Universe}, M. Kafatos, ed., Kluwer, Dortrecht, 69 (1989).
\bibitem{Ein}
A. Einstein, in {\it Albert Einstein: Philosopher-Scientist},
P.A. Schilpp, ed., Illinois: The Library of Living Philosophers,
(1949), p. 672.
\bibitem{Per}
A. Peres, J. Phys. \textbf{A24}, L175 (1991).

\end{thebibliography}
\end{document}